\documentclass[10pt,journal,compsoc]{IEEEtran}
\pdfoutput=1

\ifCLASSOPTIONcompsoc
  \usepackage[nocompress]{cite}
\else
  \usepackage{cite}
\fi
\ifCLASSINFOpdf
   \usepackage[pdftex]{graphicx}
   \usepackage{epstopdf}
   \usepackage{subfigure}
   \graphicspath{{./}}
\else
   \usepackage[dvips]{graphicx}
\fi
\usepackage{amsmath}
\usepackage{amsfonts}
\usepackage{bm}
\usepackage{subeqnarray}
\usepackage{cases}
\usepackage{algorithm}  
\usepackage{algorithmicx}  
\usepackage{algpseudocode} 

\usepackage{flushend}

\usepackage{colortbl}
\usepackage{booktabs}
\usepackage{multirow}
\usepackage{tabularx}

\begin{document}
%
\title{Leveraging AI and Intelligent Reflecting Surface for Energy-Efficient Communication in 6G IoT}
%
%
%
%

\author{Qianqian~Pan,~ 
        Jun~Wu,~\IEEEmembership{Member,~IEEE,} 
        Xi~Zheng,~\IEEEmembership{Member,~IEEE,} 
        Jianhua~Li,~\\
        Shenghong~Li,~\IEEEmembership{Senior~Member,~IEEE,}
        ~and~Athanasios~V.~Vasilakos,~\IEEEmembership{Senior~Member,~IEEE}
\IEEEcompsocitemizethanks{\IEEEcompsocthanksitem Qianqian Pan, Jun Wu, Jianhua Li, and Shenghong Li are with the School of Electronic Information and Electrical Engineering, Shanghai Jiao Tong University, Shanghai 200240, China and also with Shanghai Key Laboratory of Integrated Administration Technologies for Information Security, Shanghai 200240, China (E-mail: panqianqian@sjtu.edu.cn, junwuhn@sjtu.edu.cn, lijh888@sjtu.edu.cn, shli@sjtu.edu.cn). \quad \quad \quad \quad (Corresponding author: Jun Wu). 
\IEEEcompsocthanksitem Xi Zheng is with the Intelligent Systems Research Center, and Department of Computing, Macquarie University, Macquarie Park, NSW 2109, Australia (E-mail: james.zheng@mq.edu.au).
\IEEEcompsocthanksitem Athanasios V. Vasilakos is with the University of Technology Sydney, Australia, with the Fuzhou University, Fuzhou, China, and Lulea University of Technology, Lulea, Sweden (E-mail:th.vasilakos@gmail.com)
\protect\\
}
}

\IEEEtitleabstractindextext{%
\begin{abstract}

The ever-increasing data traffic, various delay-sensitive services, and the massive deployment of energy-limited Internet of Things (IoT) devices have brought huge challenges to the current communication networks, motivating academia and industry to move to the sixth-generation (6G) network. With the powerful capability of data transmission and processing, 6G is considered as an enabler for IoT communication with low latency and energy cost. In this paper, we propose an artificial intelligence (AI) and intelligent reflecting surface (IRS) empowered energy-efficiency communication system for 6G IoT. First, we design a smart and efficient communication architecture including the IRS-aided data transmission and the AI-driven network resource management mechanisms. Second, an energy efficiency-maximizing model under given transmission latency for 6G IoT system is formulated, which jointly optimizes the settings of all communication participants, i.e. IoT transmission power, IRS-reflection phase shift, and BS detection matrix. Third, a deep reinforcement learning (DRL) empowered network resource control and allocation scheme is proposed to solve the formulated optimization model. Based on the network and channel status, the DRL-enabled scheme facilities the energy-efficiency and low-latency communication. Finally, experimental results verified the effectiveness of our proposed communication system for 6G IoT.


\end{abstract}

}

\maketitle

\IEEEdisplaynontitleabstractindextext

%
\IEEEpeerreviewmaketitle

\IEEEraisesectionheading{\section{Introduction}\label{sec:introduction}}

%
%
%
%

\IEEEPARstart{I}{n} the recent years, billions of Internet of Things (IoT) devices are distributed everywhere and envisioned to be connected \cite{liao2020cognitive} \cite{8822503}. Driven by the various IoT-enabled latency-sensitive applications and diverse heterogeneous services, the data traffic is ever-growing at an exponential rate \cite{8672279} \cite{wu2020application} \cite{zhang2019esda}. According to the report of the International Telecommunication Union (ITU), the overall mobile data traffic will increase by $55\%$ each year in the following ten years, and reach thousands of exabytes in 2030 \cite{ITU}. These myriads of IoT users, huge amounts of valuable data, as well as energy-limited IoT devices, poses huge challenges to current communication networks. The current networks, even the fifth-generation (5G) network, cannot satisfy the increasing demands for ultra-high rate and ultra-low latency data transmission under energy limitation. The sixth-generation (6G) network is expected to provide trillion-level user access, data rate up to Tera bps, and undetectable latency with super-high efficiency \cite{tariq2020speculative} \cite{dang2020should}. 6G is considered as an enabler for massive IoT data transmission with a high rate and low energy consumption.

To mitigate data transmission issues, the most common approach is to deploy more base stations (BS)/access points (AP) \cite{moh2017dynamic}. However, this method increases energy consumption and inter interference, which is not friendly for the environment. Other efforts have been done to explore novel solutions. In the paper \cite{lin2019striking}, the authors design an IoT data collection and transmission scheme with the assistant of unmanned aerial vehicles (UAV). Although this approach alleviates the issue of IoT data transmission, the cost and energy consumption are huge, which makes them inefficient and uneconomic to be deployed on a large scale in a short period of time. The authors in paper \cite{9214878} propose an AI-based dynamic resource management scheme in industrial IoT (IIoT) to reduce long-term transmission and processing delays. However, this method only optimizes the resource of edge servers rather than the whole system, which is insufficient. Thus, it is crucially important and urgently needed to investigate the high-efficiency and low-latency communication system for 6G IoT.

\begin{figure*}[ht] 
\centering
\includegraphics[width=6.5in]{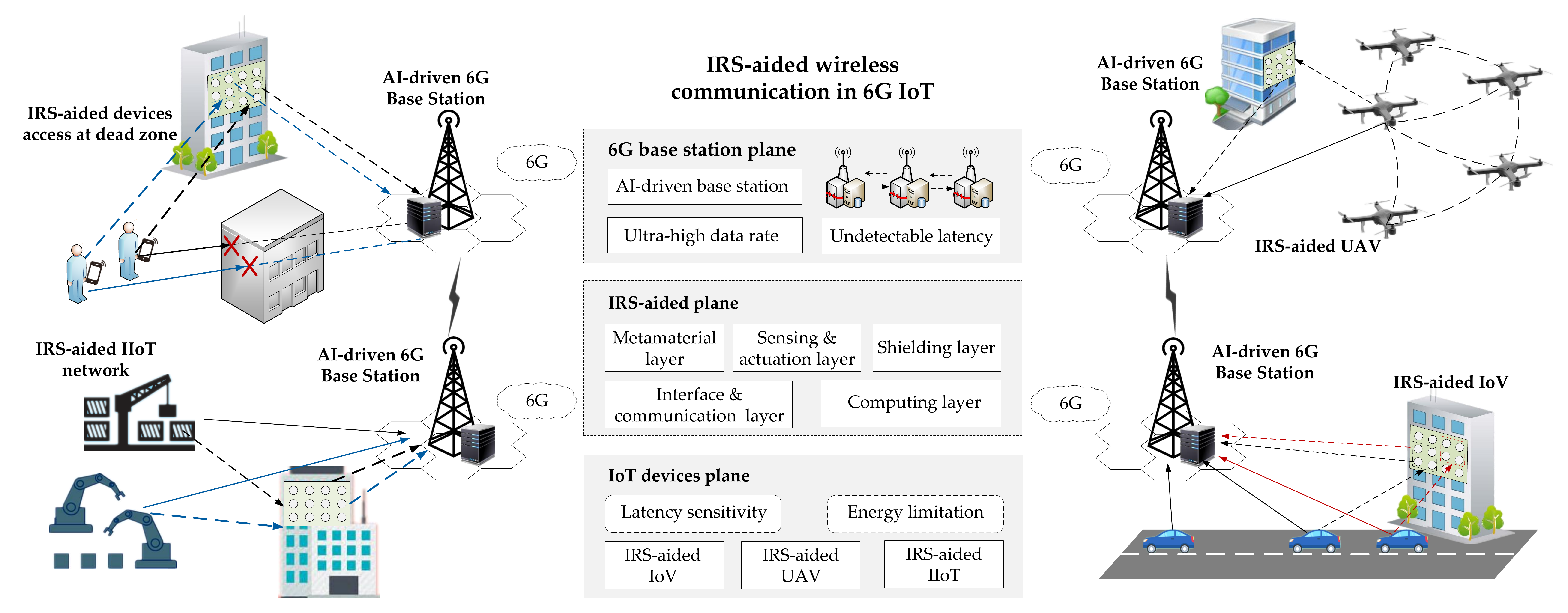}
\centering
\caption{Scenario of AI and IRS empowered communication for 6G IoT}
\label{senario}
\end{figure*}

As a promising technology in 6G, intelligent reflecting surface (IRS) can improve communication performance via modifying the signal propagation environment \cite{han2020intelligent} \cite{cai2020resource}. IRS consists of metamaterial layer, sensing \& actuation layer, shielding layer, computing layer, and interface \& communication layer \cite{huang2020holographic}. Compared with traditional technologies (e.g. BS/AP), IRS has features of nearly-passive, inexpensive, and thin, which means high energy efficiency and easy to be deployed on the walls, buildings, and ceilings \cite{wu2019towards}. These properties make it suitable and feasible for applying IRS to IoT communication system, where the capacity and energy of the devices are limited. As one of the most important trends in 6G, artificial intelligence (AI) has wide applications in various fields, e.g. image recognition, intrusion detection, and etc. \cite{sun2020drcnn} \cite{nguyen2019deep}. Unlike the conventional technologies that mainly rely on expert knowledge, AI has the learning ability to solve complex tasks intelligently and dynamically \cite{xiong2019deep}. Since the changing IoT communication environment and the strong processing capability of AI, we introduce AI to empower the data transmission in 6G IoT.

In this paper, we investigate the energy-efficiency communication system for 6G IoT by leveraging AI and IRS technologies. As shown in Fig. \ref{senario}, the IRS is deployed at the edge of the cell to assist the data transmission, while AI is equipped at the BS to provide intelligent decisions in various scenarios. Our main contributions are summarized as follows:
\begin{itemize}
  \item We propose an IRS and AI empowered communication architecture for 6G IoT to realize energy-efficiency and low-latency data transmission. In the proposed architecture, an IRS-aided mechanism is designed to assist data transmission, and an AI-driven mechanism is proposed to manage network resources.
  
  \item To improve communication performance, we design a joint optimization model to maximize the energy efficiency of the whole network under given transmission latency. In the designed model, all participants of the 6G IoT communication system including IoT devices, IRS-aided channel, and the BS are involved to jointly optimize system performance.

  \item A DRL-driven network resource control and allocation scheme is proposed. Based on the status of channel and network, the DRL-enabled scheme determines the IoT transmission power, IRS-reflection phase shift, and BS detection matrix. The scheme guilds the configurations of the whole system and facilitates energy-efficiency communication in 6G IoT.
  
\end{itemize}

The remainder of this paper is organized as follows. The related work of AI and IRS technologies in 6G is described in Section \ref{sec:related}. In Section \ref{sec:archMod}, the proposed architecture and the system model of the AI and IRS empowered communication system for 6G IoT are developed. In Section \ref{sec:DRL}, a DRL-driven network resource control and allocation scheme is implemented. Section \ref{sec:result} discusses the experiments and results. Finally, Section \ref{sec:conclusion} concludes this paper.

\section{Related Work}
\label{sec:related}

\subsection{Intelligent Reflecting Surface for 6G}
Compared with the first five generations wireless network which only operate on the transceivers to improve performance, 6G network is expected to realize the entire communication process controllable and programmable. IRS is an emerging technology to program the communication environment via numerous independently controllable reflecting elements. Extensive research works have been done to explore the benefits of IRS in 6G wireless communication \cite{pan2020intelligent} \cite{pan2020multicell}. The article \cite{wang2020channel} investigates channel estimation for IRS-assisted multi-user communications for 6G, where the framework, algorithms, and analysis are included. The authors in \cite{xu2020resource} propose the IRS-aided full-duplex cognitive radio systems, which allocates system resource and improves system performance. Authors of \cite{bai2020latency} propose an IRS-aided mobile edge computing system, in which the data off-loading and processing latency are optimized. Shafique et al. propose an integrated UAV-IRS relaying system to enhances the communication between ground users \cite{9234511}.

Since IRS is an emerging technology in 6G, the existing works of IRS for 6G are mainly limited to wireless cellular communications, cognitive radio, edge computing, and other related fields. Although the combination of IRS with AI could improve the data transmission performance in 6G IoT, it is ignored by the previous studies.

\subsection{Artificial Intelligence for 6G}
Intelligence is a native feature of 6G, which allows the network to adapt to the changing environment, deal with complex missions, and make smart decisions automatically \cite{saad2019vision} \cite{wang2020model} \cite{deng2020analysis}. Since the excellent learning ability, AI has been applied to 6G/beyond 5G (B5G) in various areas. Letaief et al. investigate the AI-empowered 6G wireless communication, where AI-enabled technologies, methodologies, and key trends for 6G design and optimization are discussed \cite{letaief2019roadmap}. The work in \cite{pan2020blockchain} proposes an AI and blockchain empowered trust-information centric network architecture for B5G, where deep reinforcement learning (DRL) is utilized to calculate the content credibility. The authors of \cite{lin2020blockchain} propose a joint power transfer and AI design for IoT to realize blockchain-based energy-knowledge trading. Gupta et al. design a secure and self-manageable telesurgery system towards 6G based on AI and blockchain, where the eXtreme Gradient Boosting (XGBoost) method is utilized to classify the diseases \cite{9286685}.

In general, the AI for 6G applications has been well studied, but most previous works ignore the utilization of AI technology to control and allocate resources of the whole network in 6G IoT, which involves IoT devices, the BS, and the channel.

\section{Proposed AI and IRS Empowered Communication System for 6G IoT}
\label{sec:archMod}
In this section, we first design architecture of the AI and IRS-driven wireless communication in 6G IoT, where the basic compositions and structure, IRS-aided data transmission, and AI-driven network resource management are described. After that, the system model is demonstrated from the channel, transmission, power consumption, and energy efficiency maximization perspectives. 

\subsection{Architecture of AI and IRS-Driven Communication}
\label{sec:architecture}

\subsubsection{Basic Composition and Structure}
\begin{figure}[t]
\centering
\includegraphics[width=3.5in]{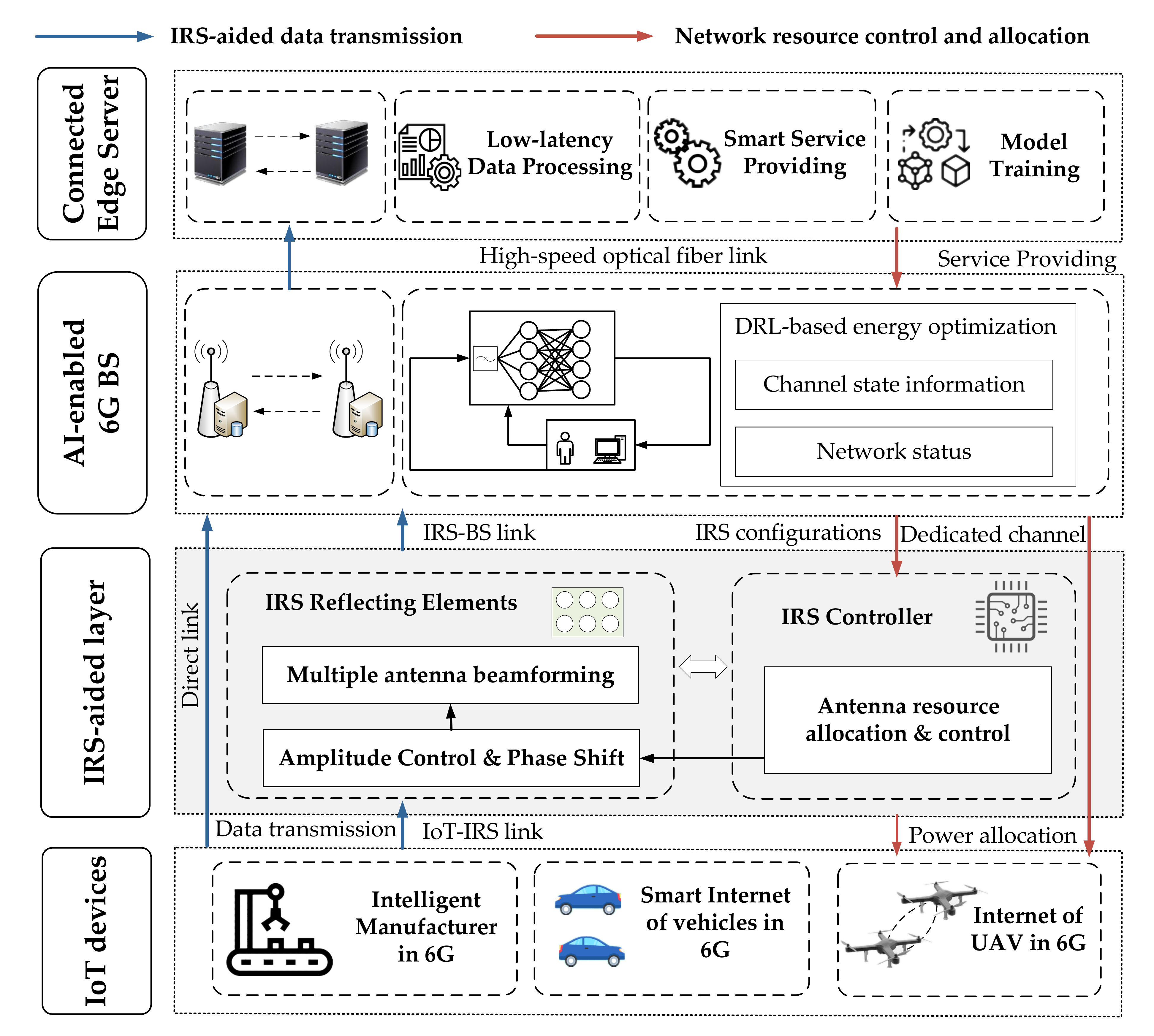}
\centering
\caption{Proposed architecture of the AI and IRS-driven wireless communication in 6G IoT }
\label{architecture}
\end{figure}

The proposed architecture is presented in a hierarchical structure, which is shown in Fig. \ref{architecture}. There are four layers of the proposed architecture, i.e. IoT devices layer, IRS-aided communication layer, AI-driven 6G BS layer, and the connected edge servers layer. The details of these four parts are illustrated as follows:

\begin{itemize}
  \item 6G IoT devices layer: This layer consists of numerous and ubiquitous underlying devices, e.g. autonomous vehicles, UAVs, and industrial equipments. These IoT devices are usually utilized to sense, collect, and even preliminarily preprocess the data. With limited energy, caching, and computing capacity, the 6G IoT devices periodically send the local data to the nearby edge server via the BS. 
  \item IRS-aided communication layer: IRS planes are deployed distractedly in each cell to assist the 6G IoT communication. The IRS consists of an IRS controller and massive IRS passive reflecting elements. The IRS controller is responsible for the management of the IRS reflecting elements. Correspondingly, under the instruction of the IRS controller, the IRS reflecting elements proactively adjust the impinging signal, thereby collaboratively changing the signal propagation.

  \item AI-driven 6G BS layer: This 6G BS layer provides access services for IoT devices, that is, the IoT users can communicate with the corresponding edge servers via the nearby BS. The 6G BS in this layer equips with constrained computing and storage capabilities, which makes it possible to integrate and preprocess the received data. Moreover, with AI function, the 6G BS can manage and configure network resources intelligently according to the dynamic network environment. 
  
  \item Connected edge server layer: The edge server is co-located with the corresponding BS, where they are connected through high-speed and high-throughput optical fibers. Data collected/generated by IoT devices is finally transmitted to the edge server via the BS. The edge server processes the data and provides service for both IoT devices and the BS.
\end{itemize}

\subsubsection{IRS-Aided Data Transmission in 6G IoT}
As shown in Fig. \ref{senario} and Fig. \ref{architecture}, there are two propagation links for the IoT devices to transmit data towards the BS, i.e. the direct link and the reflecting link. In the first propagation link, the data from IoT devices is sent to the BS through the light-of-sight (LOS) path, where signal attenuation is related to the distance between the transmitter and receiver. The direct link of data transmission may be blocked by obstacles (e.g. buildings), which is a common phenomenon in urban areas. Besides, due to the limited energy capacity of IoT devices and the path loss, the BS received signal strength from those IoT users located at the edge of the cell will be greatly decreased. All these phenomena make the signal-to-interference-plus-noise ratio (SINR) at the BS drop sharply, thereby affecting the communication rate and quality.

In the reflecting propagation mode, the IRS technology is utilized to assist data transmission from IoT to the BS. As shown in Fig. \ref{architecture}, the IRS-aided reflecting path can be divided into three parts, i.e. IoT-IRS link, IRS reflection, and IRS-BS link. First, the signal of the IoT device is sent to the IRS which usually deployed at the edge of the cell. Then, the IRS adjusts the reflection coefficient of each element to modify the amplitude and phase of the incident signal. Next, the IRS-reflection elements work collaboratively to achieve fine-grained three-dimension beamforming for directional signal transmission. Moreover, the IRS-reflected signal is added constructively/destructively with the direct-path signal to enhance/suppress the received signal at the expected direction towards the BS.

The gain of the IRS-aided 6G IoT wireless communication system includes direct gain and reflection-aided beamforming gain. The direct gain is obtained by the direct-link signal propagation, while the reflection gain is based on the assistance of IRS. These two gains increase the data transmission rate, thereby improving the performance of the whole 6G IoT communication system.

\subsubsection{AI-Driven Network Resource Management}
To manage the network resource automatically and intelligently, an AI-driven network resource management mechanism is proposed which is deployed at the BS. The designed AI-driven mechanism involves the transmitter IoT, IRS-aided channel, and the receiver BS three parties, namely all participants in the 6G IoT wireless communication system. The settings of IoT, IRS, and BS are determined by the proposed AI-driven mechanism and sent to the corresponding IoT devices and the IRS controller by dedicated channels, respectively.

According to the complex and dynamic environment of the 6G IoT communication system, DRL is utilized to design the AI-driven network resource management mechanism. As a main branch of AI, DRL can be used to optimize decision-making in a dynamic and interactive environment. Combining the reinforcement learning (RL) with deep learning, DRL has a strong ability to deal with changeable tasks. Base on the real-time channel state information (CSI) and network status, the proposed DRL-enabled resource management mechanism is utilized to improve the performance of the whole network. Specifically, the proposed mechanism jointly optimizes the transmission power of IoT devices, IRS-reflection configurations, and BS detection coefficients to maximize energy efficiency.

\subsection{AI and IRS Empowered System Model for 6G IoT}
\label{sec:model}

As shown in Fig. \ref{architecture}, we consider the AI and IRS-driven wireless communication system for 6G IoT, where $N_{I}$ single-antenna IoT devices transmit local data to the edge server through an $M_{B}$-antenna BS. In each cell, it is assumed that the edge server and the BS are co-located and connected by high-speed optical fiber. An IRS with $M_{R}$ antenna elements is placed in the cell to assist the data transmission form IoT devices to the edge server. The defined set $\mathcal{N}_{I}=\{1,2,\cdots,N_I\}$ represents the IoT users. To ease reading, the list of major notations is summarized in Table \ref{notation}.

\setlength{\extrarowheight}{0.11mm}
\begin{table}[htbp]
\renewcommand{\arraystretch}{1.2}
\caption{Major notations}
\label{notation}
\centering
\begin{tabular}{cc}

    \toprule
    Notations & Description\\
    \midrule
     $\bm{h}_{d,k}$ & direct channel between BS and the $k$th IoT device\\
     $\bm{h}_{r,k}$ & reflecting channel between IRS and the $k$th IoT device \\
     $\bm{G}$  & the channel coefficients between IRS and BS \\
     $\bm{\theta}$ & the phase shift coefficient of IRS  \\
     $N_I$  & the number of IoT devices in the cell \\
     $M_R$  & the number of IRS reflection elements  \\
     $M_B$  & the number of antennas at the BS \\
     $\eta$ & the exponent coefficient of path loss \\
     $\rho_{d},\rho_{r}$ & pass loss of the direct and reflecting link \\
     $\bm{x},\bm{y}$ & the transmitted signal and the received signal\\ 
     $\bm{p}$ & the transmission power vector for IoT devices \\
     $\bm{W}$ &  the matrix of linear multi-users detection at the BS  \\ 
     $\gamma_k$ & the SINR for the $k$th IoT device \\
     $R_k$ & the transmission rate for the $k$th IoT device \\
     $\tau_k$ & transmission delay for the $k$th IoT device\\
     $L_k$ & amount of data to be transferred of the $k$th IoT device\\
     $T_k$ & latency requirement of the $k$th IoT device\\
     $P_k$ & maximum transmission power of the $k$th IoT device\\
     $\mathcal{CN}(0,\sigma^2)$ &circularly symmetric complex Gaussian distribution\\
    \bottomrule
\end{tabular} 
\end{table}

\subsubsection{Channel Model for the AI and IRS Empowered Communication System}
In the proposed AI and IRS empowered system, an IoT device can send data to the BS by direct transmission and the IRS-aided reflection. In this paper, we assume the CSI of all the links can be estimated precisely, and these channels remain near-constant when IoT devices transmitting local data. The equivalent baseband channel between the $k$-th IoT device and the BS is denoted as $\bm{h}_{d,k} \in \mathbb{C}^{M_{B}\times 1}$, which is given by
\begin{equation}
	\begin{split}
		\bm{h}_{d,k} &= [1,e^{-j\frac{2\pi \Delta_{M_{Bx}}}{\lambda_c}} \sin \alpha_{k}^{IB} \cos \beta_{k}^{IB}, \cdots,\\
		& e^{-j\frac{2\pi \Delta_{M_{Bx}}}{\lambda_c}} (M_{Bx}-1)\sin \alpha_{k}^{IB} \cos \beta_{k}^{IB}]^{H}\\
		&\otimes  [1,e^{-j\frac{2\pi \Delta_{M_{By}}}{\lambda_c}} \sin \alpha_{k}^{IB} \sin \beta_{k}^{IB},\cdots, \\
		& e^{-j\frac{2\pi \Delta_{M_{By}}}{\lambda_c}} (M_{By}-1)\sin \alpha_{k}^{IB} \sin \beta_{k}^{IB}]^{H}\\
	\end{split}
\end{equation} 
where $\lambda_c$ is the wavelength of the center carrier. $\Delta_{M_{Bx}}$ and $\Delta_{M_{By}}$ denote the antenna separation of the BS in $x$ and $y$ dimension respectively, and $M_{B_x} \times M_{B_y} = M_B$. $\alpha_{k}^{IB}$ and $\beta_{k}^{IB}$ represent the vertical and horizontal angle from the $k$th IoT user to the BS respectively. $\otimes$ is the Kronecker product.

The IRS-reflection channel can be presented as the concatenation of from IoT device to IRS, IRS reflection, and from IRS to the BS (i.e. IoT-IRS-BS). For the $k$th IoT device, the equivalent baseband channel to the IRS is denoted as $\bm{h}_{r,k} \in \mathbb{C}^{M_{R}\times 1}$, which is given as follows:
\begin{equation}
	\begin{split}
		\bm{h}_{r,k} &= [1,e^{-j\frac{2\pi \Delta_{M_{Rx}}}{\lambda_c}} \sin \alpha_{k}^{IR} \cos \beta_{k}^{IR}, \cdots,\\
		& e^{-j\frac{2\pi \Delta_{M_{Rx}}}{\lambda_c}} (M_{Rx}-1)\sin \alpha_{k}^{IR} \cos \beta_{k}^{IR}]^{H}\\
		&\otimes  [1,e^{-j\frac{2\pi \Delta_{M_{Ry}}}{\lambda_c}} \sin \alpha_{k}^{IR} \sin \beta_{k}^{IR},\cdots, \\
		& e^{-j\frac{2\pi \Delta_{M_{Ry}}}{\lambda_c}} (M_{Ry}-1)\sin \alpha_{k}^{IR} \sin \beta_{k}^{IR}]^{H}\\
	\end{split}
\end{equation} 
where $\Delta_{M_{Rx}}$ and $\Delta_{M_{Ry}}$ denote the element separation of the IRS in $x$ and $y$ dimension respectively, and $M_{Rx}\times M_{Ry}=M_R$. Typically, the inter-distance of the IRS unit element is the order of the wavelength (i.e. half of the wavelength) or smaller than the wavelength (i.e. 5-10 times smaller than the wavelength) \cite{di2020smart}. $\alpha_{k}^{IR}$ and $\beta_{k}^{IR}$ represent the vertical and horizontal angle from the $k$th IoT user to the IRS respectively. 

The equivalent baseband channel between the IRS and the BS is denoted as $\bm{G}\in \mathbb{C}^{M_{B}\times M_R}$, which is formulated as follows:
\begin{equation}
	\begin{split}
		\bm{G} &= [1,e^{-j\frac{2\pi \Delta_{M_{Bx}}}{\lambda_c}} \sin \alpha^{BR} \cos \beta^{BR}, \cdots,\\
		& e^{-j\frac{2\pi \Delta_{M_{Bx}}}{\lambda_c}} (M_{Bx}-1)\sin \alpha^{BR} \cos \beta^{BR}]^{H}\\
		&\otimes  [1,e^{-j\frac{2\pi \Delta_{M_{By}}}{\lambda_c}} \sin \alpha^{BR} \sin \beta^{BR},\cdots, \\
		& e^{-j\frac{2\pi \Delta_{M_{By}}}{\lambda_c}} (M_{By}-1)\sin \alpha^{BR} \sin \beta^{BR}]^{H}\\
		&\otimes [1,e^{-j\frac{2\pi \Delta_{M_{Rx}}}{\lambda_c}} \sin \alpha^{RB} \cos \beta^{RB}, \cdots,\\
		& e^{-j\frac{2\pi \Delta_{M_{Rx}}}{\lambda_c}} (M_{Rx}-1)\sin \alpha^{RB} \cos \beta^{RB}]\\
		&\otimes  [1,e^{-j\frac{2\pi \Delta_{M_{Ry}}}{\lambda_c}} \sin \alpha^{RB} \sin \beta^{RB},\cdots, \\
		& e^{-j\frac{2\pi \Delta_{M_{Ry}}}{\lambda_c}} (M_{Ry}-1)\sin \alpha^{RB} \sin \beta^{BR}]\\
	\end{split}
\end{equation} 
where $\alpha_{k}^{BR}$ and $\beta_{k}^{BR}$ denote the vertical and horizontal angle from the BS to the IRS respectively. $\alpha_{k}^{RB}$ and $\beta_{k}^{RB}$ are the vertical and horizontal angle from the IRS to the BS respectively. It can be obtained that $\alpha_{k}^{BR}=\alpha_{k}^{RB}$ and $\beta_{k}^{BR}=\beta_{k}^{RB}$.

In this paper, the amplitude reflecting coefficient of IRS is set as 1. The phase shift coefficient of IRS is denoted as $\bm{\theta}=[\theta_1,\theta_2,\cdots,\theta_{M_R}]^{T}$, where $\theta_i \in [0,2\pi)$ for $i \in \{1,2,\cdots,M_R\}$. The reflection-coefficient matrix of IRS is presented as $\mathbf{\Theta}=\mathrm{diag}\{e^{j\theta_1},e^{j\theta_2},\cdots,e^{j\theta_{M_R}}\}$. The IRS phase shift is calculated by the DRL-enabled agent at the BS according to the dynamic channel and IRS states. Then, the phase shift settings are sent to the IRS controller via the dedicated channel. The equivalent baseband channel of IoT-IRS-BS for the $k$th IoT device can be modeled as $\bm{G} \bm{\Theta} \bm{h}_{r,k}$.

Due to the reflection of the IRS, the channel coefficient of the proposed system from the $k$th IoT device to the BS is modeled as follows:
\begin{equation}
	\begin{split}
		\bm{h}_{k} = \sqrt{\rho_{d,k}} \cdot \bm{h}_{d,k} +  \sqrt{\rho_{r,k}} \cdot \bm{G}\bm{\Theta}\bm{h}_{r,k}
	\end{split}
\end{equation} 
where $\rho_{d,k}$ is the path loss of the direct link between the $k$th IoT device and the BS, and $\rho_{r,k}$ is the path loss of the reflecting link from IoT device $k$ to the BS via the IRS. The path loss $\rho_{d,k}$ is presented as $\rho_{d,k}=l_0/d(S_{I,k},S_{B})^{\eta}$, where $S_{I,k}$ and $S_{B}$ are the position vectors of the $k$th IoT device and the BS respectively. $d(a,b)=||a-b||_2$ denotes the distance between the two input vectors. $\eta$ is the exponent coefficient of path loss. $l_0$ is the path loss of the unit reference distance. For the reflecting link IoT-IRS-BS, the pass loss is given as follows:
\begin{equation}
	\begin{split}
		\rho_{r,k} = \frac{l_0}{ ( d(S_{I,k},S_{R}) +d(S_{R},S_{B}) )^{\eta} }
	\end{split}
\end{equation} 
where $S_{R}$ represents the position vector of the IRS. 

\subsubsection{Transmission Model of the AI and IRS-Driven System}

In the proposed system, the $N_I$ IoT devices generate and collect local data, and then transmit the data to the edge server via BS through the direct and reflecting links. For the $k$th IoT device, the tuple $(L_k, T_k, P_k)$ is utilized to denote its total number of bits to be transmitted, its transmission delay requirements, and its maximum transmission power. 

It is assumed that the $N_I$ IoT devices transmit their local data on a given frequency band $B$ at the same time slot. The transmitting signal of the $N_I$ devices is denoted as $\bm{x}=[x_1,x_2,\cdots,x_{N_I}]^T$ and the transmission power vector is $\bm{p}=[p_1,p_2,\cdot,p_{N_I}]^T$. For the $k$th IoT device, its transmission power can be written as $p_k=\mathbb{E}(|x_k|^2) \leq P_k$. The received signal $\bm{y}\in \mathbb{C}^{M_B \times 1}$ at the BS is formulated as
\begin{equation}
	\begin{split}
		\bm{y} &= \sum\nolimits_{k=1}^{N_I} \bm{h}_k \cdot x_k + \bm{n}\\
		           &= \sum\nolimits_{k=1}^{N_I} (\sqrt{\rho_{d,k}} \cdot \bm{h}_{d,k} +  \sqrt{\rho_{r,k}} \cdot \bm{G}\bm{\Theta}\bm{h}_{r,k}) \cdot x_k + \bm{n}\\
	\end{split}
\end{equation} 
where $\bm{n}=[n_1,n_2,\cdots,n_{M_B}]^T$ represents the noise vector at the BS. We assume $n_m \sim \mathcal{CN}(0,\sigma^2)$ for $m=1,2,\cdots, M_B$, where $\mathcal{CN}(0,\sigma^2)$ is the circular symmetric complex Gaussian distribution with zero mean and $\sigma^2$ variance. It is assumed the BS performs the linear multi-users detection to estimate the signal transmitted by IoT devices. Accordingly, the estimated signal $\hat{\bm{x}}$ is given as
\begin{equation}
	\begin{split}
		\hat{\bm{x}}=\bm{W}^H\bm{y} = \bm{W}^H \cdot (\sum\nolimits_{k=1}^{N_u} \bm{h}_k \cdot x_k + \bm{n})
	\end{split}
\end{equation} 
where $\bm{W}=[\bm{w}_1, \bm{w}_2,\cdots, \bm{w}_{N_I}] \in \mathbb{C}^{M_B \times N_I}$ is the matrix of linear multi-users detection. For the $k$th IoT device in the system, the estimation of the transmitted signal is written as follows:
\begin{equation}
	\begin{split}
		\hat{x}_k= \bm{w}_k^H [ \sum\nolimits_{k=1}^{N_I} (\sqrt{\rho_{d,k}} \cdot \bm{h}_{d,k} +  \sqrt{\rho_{r,k}} \cdot \bm{G}\bm{\Theta}\bm{h}_{r,k}) \cdot x_k + \bm{n}]
	\end{split}
\end{equation} 

The SINR of the received signal for the $k$th device is obtained as follows:
\begin{equation}
	\begin{split}
	&\gamma_k(\bm{w}_k,\bm{\theta},\bm{p})\\
	&=\frac{p_k|\bm{w}_k^H (\sqrt{\rho_{d,k}}  \bm{h}_{d,k} +  \sqrt{\rho_{r,k}}  \bm{G}\bm{\Theta}\bm{h}_{r,k}) |^2}{\sum \nolimits_{i=1, i\neq k}^{N_I} p_i| \bm{w}_k^H (\sqrt{\rho_{d,i}}  \bm{h}_{d,i} +  \sqrt{\rho_{r,i}}  \bm{G}\bm{\Theta}\bm{h}_{r,i}) |^2 + \sigma^2|\bm{w}_k^H|^2}
  \end{split}
\end{equation} 
In this paper, it is assumed that the perfect transmission scheme is utilized. Thus, we can obtain the maximum achievable transmission rate of the $k$th IoT device as follows:
\begin{equation}
	\begin{split}
		R_k(\bm{w}_k,\bm{\theta},\bm{p}) = B\log_2 [1+\gamma_k(\bm{w}_k,\bm{\theta},\bm{p})]
	\end{split}
\end{equation} 
where the unit of $R_k$ is bits per second (bps). Then, the signal transmission delay $\tau_k$ can be formulated as
\begin{equation}
	\begin{split}
		\tau_k = \frac{L_k}{R_k(\bm{w}_k,\bm{\theta},\bm{p})}=\frac{L_k}{B\log_2 [1+\gamma_k(\bm{w}_k,\bm{\theta},\bm{p})]}
	\end{split}
\end{equation}

\subsubsection{Power Consumption Model of the Proposed System}
The power consumption of the proposed AI and IRS-driven communication system for 6G IoT is consists of the following three parts:
\begin{itemize}
  \item Signal transmission power consumption on IoT devices: The power cost for the local data transmission from the IoT is decided by all IoT devices' signal sending power, which is given as $P_{IoT} = \sum \nolimits_{k=1}^{N_I} p_k$.
  \item Reflecting power consumption on IRS: As an energy-limited device, the IRS consumes energy to control the phase shift of the signal. The total power consumption of the IRS during signal transmission is formulated as $P_{IRS} = M_R \cdot p_R$, where $p_R$ represents the power cost of each IRS element for signal phase shift.
  \item Signal processing power consumption on BS: When the signal arrives, the BS detects and preprocesses the received data, and sends them to the connected edge server via high-speed optical fiber. The circuit power consumption by the modules at the BS is given as $P_{BS}$.
\end{itemize}
Therefore, the total power consumption of the AI and IRS empowered communication system for 6G IoT is formulated as follows:
\begin{equation}
		P_{total}(\bm{p}) = P_{IoT}+P_{IRS}+P_{BS}=\sum \nolimits_{k=1}^{N_I} p_k+M_R \cdot p_R + P_{BS}
\end{equation}

\subsubsection{Energy Efficiency Maximization Model}
In the 6G IoT system, due to the growth of energy-limited IoT devices, ultra-low latency requirements for communication, and dynamic network status, it is important to improve energy efficiency and reduce time cost. For the proposed AI and IRS-driven 6G IoT communication system, we jointly optimize the resource of IoT devices, the BS, and the IRS to improve the energy efficiency of the whole system under the given latency requirements. Thus, we formulate the model as maximizing the system energy efficiency subject to the phase shift restrictions, transmit power constraints, and maximal latency requirements. The optimization model is expressed by
	\begin{align}
		 &\max \limits_{\bm{W},\bm{\theta},\bm{p}} \frac{\sum \nolimits_{k=1}^{N_I}R_k(\bm{w}_k,\bm{\theta},\bm{p})}{P_{total}(\bm{p})} \label{op}\\
		&s.t. ~~0\leq \theta_i \leq 2\pi, i=1,2,\cdots, M_R, \tag{\ref{op}{a}} \label{opa}\\
		     &~~~~~~~0\leq p_k \leq P_k, k=1,2,\cdots, N_I, \tag{\ref{op}{b}} \label{opb}\\
		     &~~~~~~~\tau_k \leq T_k, k=1,2,\cdots, N_I. \tag{\ref{op}{c}} \label{opc}
	\end{align}
where energy efficiency is defined as the whole transmission rate per unit energy (bps/W). In equation(\ref{op}), we formulate the optimization model with a total of three variables, i.e. transmission power of IoT devices $\bm{p}$, IRS phase shift $\bm{\theta}$, and BS multi-user detection matrix $\bm{W}$. Constraint (\ref{opa}) specifies the range of the IRS element phase shift. (\ref{opb}) ensures the power consumption of each IoT device does not exceed the its maximum transmission power. Finally, constraint (\ref{opc}) guarantees the latency requirements of IoT devices.

\section{DRL-Driven Network Communication Resource Control and Allocation}
\label{sec:DRL}
In this section, a novel DRL-driven scheme used for network resource control and allocation is proposed. We first introduce the framework of the proposed DRL-based scheme. Then, the description of the DRL-enabled resource control and allocation algorithm is presented.

\begin{figure*}[!t]
\centering
\includegraphics[width=6.5in]{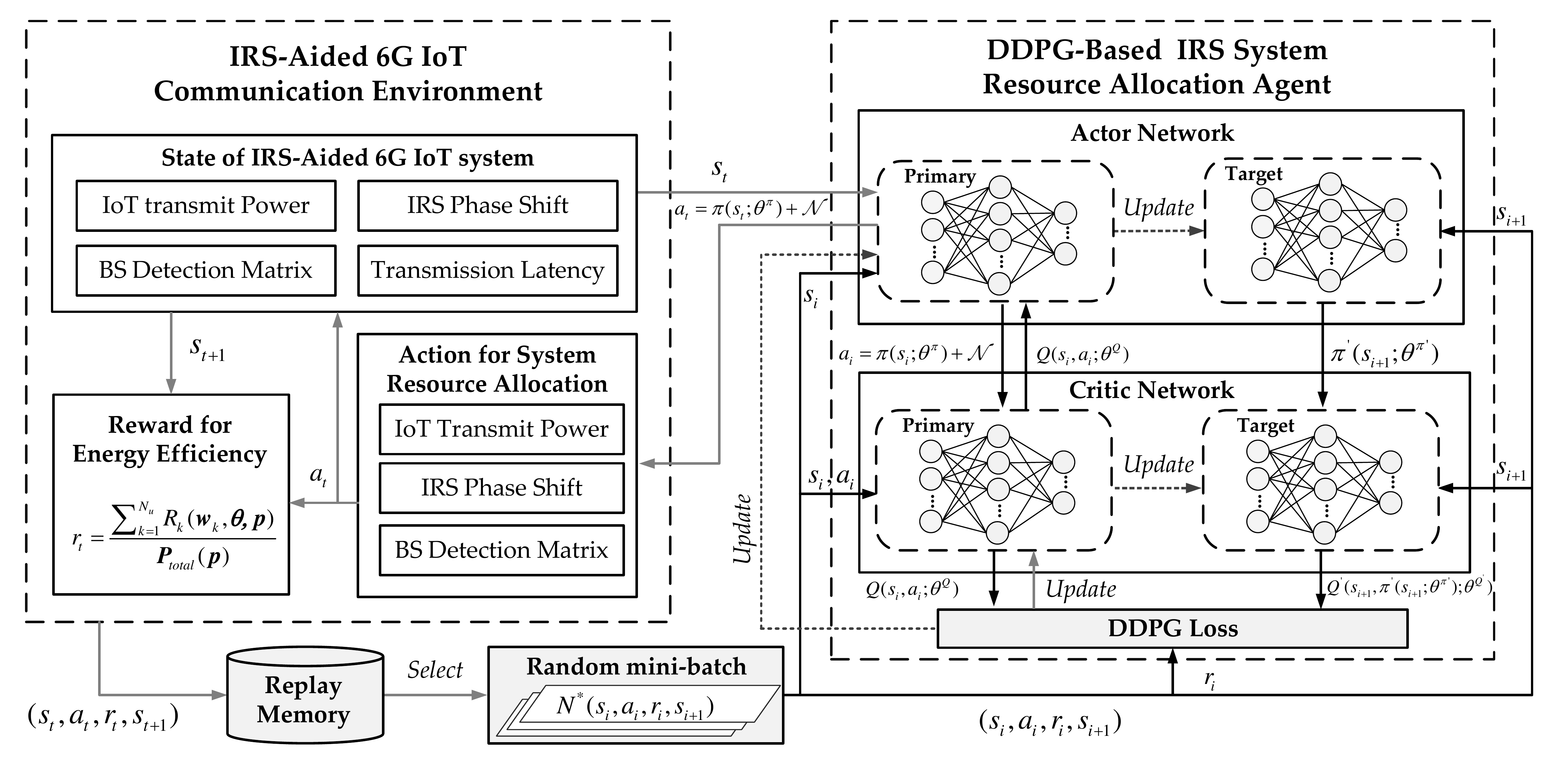}
\centering
\caption{Framework of DRL empowered communication network resource control and allocation}
\label{DRL}
\end{figure*}

\subsection{Framework of DRL-Based Resource Control and Allocation}

The proposed framework for DRL-based resource control and allocation is shown in Fig. \ref{DRL}. In our proposed architecture, the BS of each cell is regarded as the DRL agent. The environment is the whole IoT communication network, including the BS, IRS-aided channel, and IoT devices. Our goal of the proposed DRL-based scheme is to jointly optimize the IoT devices' transmission power vector $\bm{p}$, IRS-reflection phase shift $\bm{\theta}$, and the BS detection matrix $\bm{W}$ to maximize network energy efficiency under given latency limitations. CSI and 6G IoT network status are collected and sent to the DRL agent. Then, the agent takes action to the IRS-aided 6G IoT communication environment and obtain the corresponding reward from the environment. The details of DRL state, action, and reward are described as follows:  

\begin{itemize}
  \item State: The state of the AI and IRS-enabled 6G IoT communication environment consists of four parts, i.e. the IoT transmission power $\bm{p}=[p_1,p_2,\cdots,p_{N_I}]^T$, IRS-reflection phase shift $\bm{\theta}=[\theta_1,\theta_2,\cdots,\theta_{M_R}]^T$, BS detection matrix $\bm{W}=[\bm{w}_1,\bm{w}_2,\cdots,\bm{w}_{M_B}]$, and transmission latency $\bm{\tau}=[\tau_1,\tau_2,\cdots,\tau_{N_I}]^T$. We use $s_t=[\bm{p}^{(t)},\bm{\theta}^{(t)},\bm{W}^{(t)},\bm{\tau}^{(t)}]$ to denote the state of the agent at the iteration $t$.
  \item Action: The action of the agent in our proposed DRL-based scheme consists of the IoT transmission power $\bm{p}$, IRS-reflection phase shift $\bm{\theta}$, and the BS detection matrix $\bm{W}$. Hence, the action is expressed as $a_t=[\bm{p}^{(t)},\bm{\theta}^{(t)},\bm{W}^{(t)}]$. At the iteration $t$, the agent at the BS makes action decision $a_t$ towards the environment state $s_t$.
\item Reward: In this proposed DRL-enabled scheme, the definition of the reward consists of two parts. One is related to the system energy efficiency; another is the penalty if the transmission time exceeds the latency limitation. The reward of the proposed scheme is formulated as follows:
\begin{equation}
    r_t = \frac{\sum \nolimits_{k=1}^{N_I}R_k^{(t)}[\bm{w}_k^{(t)},\bm{\theta}^{(t)},\bm{p}^{(t)}]}{P_{total}^{(t)}(\bm{p}^{(t)})} - \sum \nolimits_{k=1}^{N_I} s_{k,t}\zeta_{k}
\end{equation} 
where the first term is the energy efficiency and the second term is the penalty. $s_{k,t}=1$ when the transmission time of the $k$th IoT device exceeds its delay requirement. Otherwise, $s_{k,t}=0$. $\zeta_{k}$ is the penalty factor. The penalty term can drive the communication system to avoid actions which leads to high latency. In the learning process of the DRL-based scheme, the weights are modified to optimize the defined reward function.
\end{itemize}

The goal of the proposed DRL-based scheme is to find the best policy mapping system state $s_t$ to $a_t$ that maximizes the cumulative discounted reward, which is given as
\begin{equation}
    G_t = \sum_{i=t}^{T} \gamma^{(i-t)} r_{i}
\end{equation} 
where $\gamma$ is the discount factor and $0\leq \gamma \leq 1$, and $T$ is the final timeslot.

\subsection{DRL-Enabled Resource Control and Allocation Algorithm}
Since the state and action variables of the DRL-based resource control and allocation scheme are continuous, deep deterministic policy gradient (DDPG) algorithm is utilized to solve the optimization model formulated in Eq. (\ref{op}). As is shown in Fig. \ref{DRL}, the DDPG is the policy-based actor-critic structure and use deep neural networks to fit the strategy and value function. Both actor and critic networks consist of primary and target networks, where target networks are the old version for corresponding primary networks.

In the learning process, the BS agent determines the action of network resource allocation $a_t$ according to the system state $s_t$ by leveraging the primary actor network $\mu(s|\theta^{\mu})$. The action can be given as
\begin{equation}
    a_t = \mu(s_t|\theta^{\mu}) + \mathcal{N}
\end{equation} 
where $\mathcal{N}$ is the random noise added in action exploration. $\theta^{\mu}$ is parameter of primary actor network. The IRS-aided 6G IoT communication environment runs the action $a_t$ and calculates the reward $r_t$ as well as the next state $s_{t+1}$. Next, the tuple $(s_t,a_t,r_t,s_{t+1})$ is stored in the replay memory with the size of $M_{RM}$, which is deployed at the BS. During the training process of the proposed DRL-based scheme, a mini-batch with $N (N\ll M_{RM})$ tuples, denoted as $N*(s_i,a_i,r_i,s_{i+1}), i\in[1,2,\cdots,N]$, is selected randomly from the replay memory to train the DDPG networks.

We update the parameter $\theta^{Q}$ of the primary critic network $Q(s,a|\theta^{Q})$ as 
\begin{equation}
    \theta^{Q} = \theta^{Q} + \alpha_Q \mathbb{E}[2(z_i-Q(s_i,a_i|\theta^{Q})) \nabla_{\theta^Q} Q(s,a|\theta^{Q})|_{s=s_i,a=a_i}]
    \label{updatecritic}
\end{equation}
where $z_i$ is the target value of the proposed DRL-enabled resource control and allocation scheme, and is related to the target actor network $\mu'(s|\theta^{\mu'})$ and the target critic network $Q'(s,a|\theta^{Q'})$
\begin{equation}
    z_i = r_i + \gamma Q'(s_{i+1},\mu'(s_{i+1}|\theta^{\mu'})|\theta^{Q'})
    \label{eta}
\end{equation}

In the proposed DRL-based scheme, the weight of the primary actor network $\theta^{\mu}$ is updated based on the policy gradient method \cite{sutton1999policy}, given as
\begin{equation}
    \theta^{\mu} = \theta^{\mu} + \alpha_\mu \mathbb{E}[ \nabla_a Q(s,a|\theta^{Q})|_{s=s_i,a=\mu(s_i|\theta^{\mu})} \nabla_{\theta^{\mu}} \mu(s|\theta^{\mu})|_{s=s_i}]
    \label{updateactor}
\end{equation}

As the old version of the primary network, the parameters of target critic network $\theta^{Q'}$ and actor network $\theta^{\mu'}$ at the 6G BS are updated as follows:
\begin{equation}
    \begin{split}
    &\theta^{Q'} = \varepsilon \theta^{Q} + (1-\varepsilon) \theta^{Q'}\\
    &\theta^{\mu'} = \varepsilon \theta^{\mu} + (1-\varepsilon) \theta^{\mu'}\\
    \end{split}
    \label{target}
\end{equation}
where $\varepsilon$ is the update rate.

The proposed DRL empowered resource control and allocation algorithm for 6G IoT system is presented in Algorithm \ref{algorithm}. 

\makeatletter  
\def\BState{\State\hskip-\ALG@thistlm}  
\makeatother 
\begin{algorithm}[h] 
\caption{DRL empowered resource control and allocation algorithm for 6G IoT system} 	
\label{algorithm}
\begin{algorithmic}[1] 
\State Initialize primary networks $\mu(s|\theta^{\mu})$ and $Q(s,a|\theta^{Q})$
\State Initialize target networks $\mu'(s|\theta^{\mu'})$, and $Q'(s,a|\theta^{Q'})$:
\State \quad \quad $\theta^{\mu'} \leftarrow \theta^{\mu}$,  $\theta^{Q'} \leftarrow \theta^{Q}$
\State Initialize the replay memory $R_{mem} $ with $M_{RM}$ size
   \For{episode $e$ $\leftarrow 1$ to $E$}
   		\State $s_1 \leftarrow$ observed current state of the IoT environment
   		\For{slot $t \leftarrow 1$ to $T$}
   			 \State Select the resource control action $a_t \leftarrow \mu(s_t|\theta^{\mu})$
   			 \State Execute $a_t$ and obtain $r_t$ and $s_{t+1}$ 
   			 \State Store $(s_t,a_t,r_t,s_{t+1})$ to $R_{mem}$
   			 \State Select randomly $N^*(s_i,a_i,r_i,s_{i+1})$ from $R_{mem}$
   			 \For{ $i \leftarrow 1$ to $N$}
	   			 \State Calculate the target value based on Eq. (\ref{eta}):
   			 	 \State \quad $z_i \leftarrow r_i + \gamma Q'(s_{i+1},\mu'(s_{i+1}|\theta^{\mu'})|\theta^{Q'})$
             \EndFor
   			 \State Update primary networks based on Eq. (\ref{updatecritic})(\ref{updateactor}):
   			 \State \quad \quad$\theta^{Q} \leftarrow UpdateCriticParameters() $ 
   			 \State \quad \quad $\theta^{\mu} \leftarrow UpdateActorParameters()  $ 
   			 \State Update target networks based on Eq. (\ref{target})
   			 \State \quad \quad $\theta^{Q'} \leftarrow \varepsilon \theta^{Q} + (1-\varepsilon) \theta^{Q'}$
   			 \State \quad \quad $\theta^{\mu'} \leftarrow \varepsilon \theta^{\mu} + (1-\varepsilon) \theta^{\mu'}$
   		\EndFor
   \EndFor

\end{algorithmic} 
\label{code} 
\end{algorithm}

\setlength{\extrarowheight}{0.1mm}
\begin{table}[htbp]
\renewcommand{\arraystretch}{1.2}
\caption{Parameter settings}
\label{patameter}
\centering
\begin{tabular}{lr}
    \toprule
    {Description} & {Parameter and Value}\\
    \toprule
     \multirow {2}{*}{ Position information}  & $R_{BS}=100m$ \\
                                                             & $S_{BS}=(0,0),S_{IRS}=(R_{BS},0)$ \\
     \midrule
     \multirow {6}{*}{ Communication settings} & $B=1$MHz \\
     		 & $\eta_{d}=3.5, \eta_{r}=2.2$ \\
      	 	 & $l_0=30$dBm \\
     		 & $\sigma^2=-114$dBm\\
		     & $P_k=5$dBm\\
		     & $P_{BS}=30$dBm, $p_R=6.5$dBm \\
     
     \hline
     \multirow {2}{*}{ Transmission model}  & $T_{k}=8$s \\
                                                                  & $L_{k}\in[250,350]$ kb \\
     \hline
     \multirow {2}{*}{ DRL coefficients }  & $\zeta_{k}=1$\\
                                                             & $R_{RM}=10^6$, $N=64$\\
    \toprule
\end{tabular} 
\end{table}

\section{Empirical Study}
\label{sec:result}
In this section, a series of experiments are implemented to verify the effectiveness of the proposed AI and IRS empowered energy-efficiency communication system. The experiment setup is discussed in subsection \ref{settings}. Then, the comparison analysis with the baseline method and the impact of parameter settings on the mechanism is analyzed in subsection \ref{other} and \ref{results}, respectively.

\subsection{Experiment Setup}
\label{settings}

A signal-cell IoT data transmission system with multiple IoT users is implemented in this section. The coverage radius of the BS is $R_{BS}=100m$ and the position of the BS is considered as $S_{BS}=(0,0)$. The IRS is deployed at the edge of the cell, whose position can be presented as $S_{IRS}=(R_{BS},0)$. The IoT devices are distributed randomly in the coverage of the BS. For all links involved in this paper (i.e. IoT-BS link, IoT-IRS link, and IRS-BS link), we consider them as independent Rayleigh fading channels which are distributed as $\bm{h}_{d,k}\sim \mathcal{CN}(\bm{0},\mathbf{I})$, $\bm{h}_{r,k} \sim \mathcal{CN}(\bm{0},\mathbf{I})$, and $g_{i,j} \sim \mathcal{CN}(0,1)$, for $k=[1,2,\cdots,N_I]$, $i=[1,2,\cdots,M_B]$, and $j=[1,2,\cdots,M_R]$. The bandwidth for IoT data transmission is set as $B=1$MHz. The pass loss of unit reference distance is $l_0=30$dBm, and the exponent coefficient for direct link and IRS-reflection link are $\eta_d=3.5$ and $\eta_r=2.2$, respectively. We set the variance of the noise at the BS as $\sigma^2=-114$dBm, and consider the maximum transmission power of IoT devices as $P_k=5$dBm, the power consumption of the BS as $P_{BS}=30$dBm, and the energy cost of one IRS-reflection element as $p_R=6.5$dBm. We assume the amount of data that needs to be transmitted of IoT users distributes uniformly at $[250,350]$Kb. Other settings including $T_k$, $\zeta_k$, $R_{RM}$, and $N$ are given in the Table \ref{patameter}.

\begin{figure}[tbp]
\centering
\includegraphics[width=3.3in]{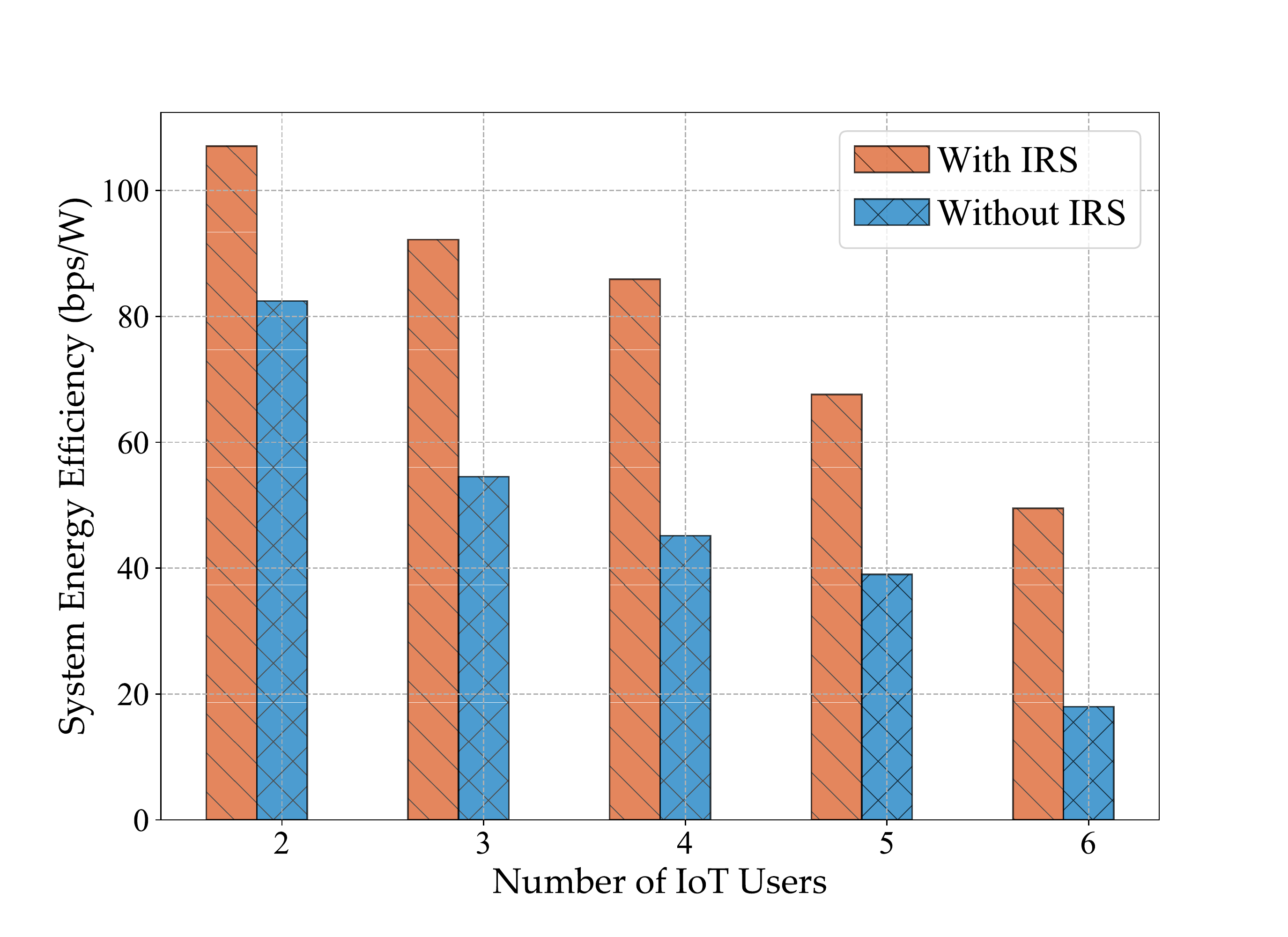}
\centering
\caption{Experimental results of the energy efficiency versus the number of IoT devices}
\label{EE-IoT}
\end{figure}

\begin{figure}[tbp]
\centering
\includegraphics[width=3.3in]{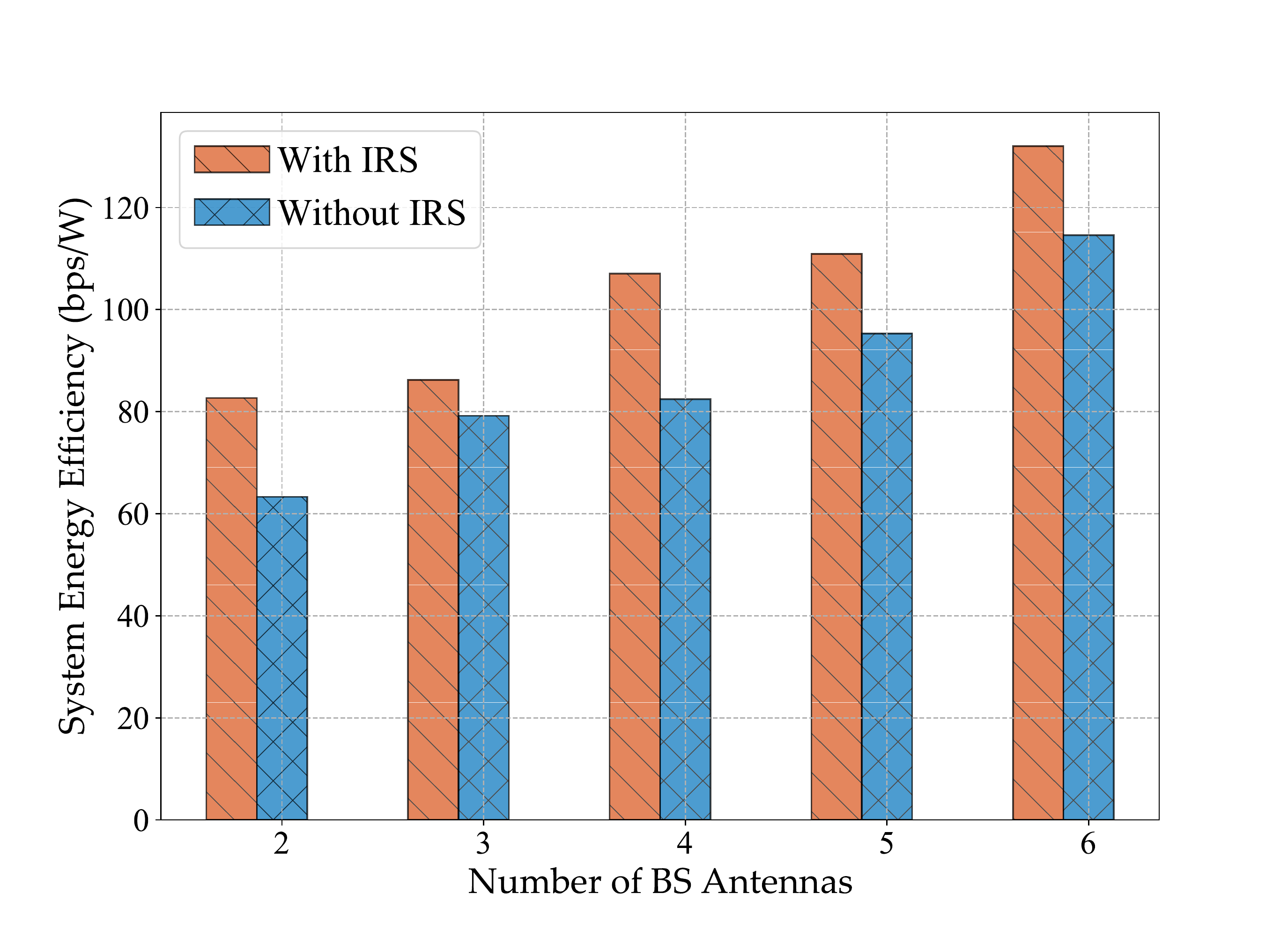}
\centering
\caption{Experimental results of the energy efficiency versus the number of BS antennas}
\label{EE-BS}
\end{figure}



\begin{figure*}[t]
\centering
\subfigure[]{
\begin{minipage}[t]{0.33\linewidth}
\centering
\includegraphics[width=2.5in]{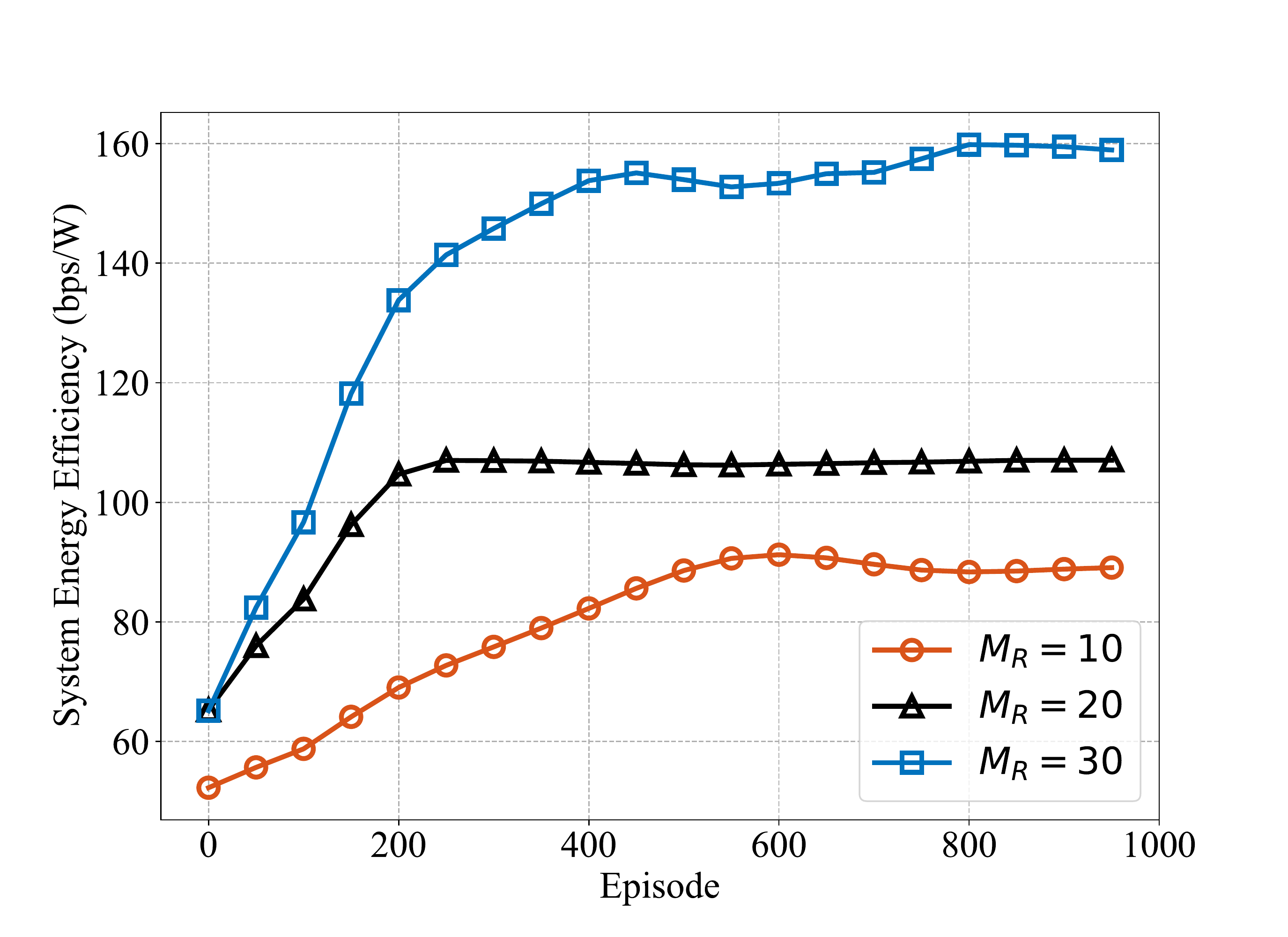}
\end{minipage}%
}%
\subfigure[]{
\begin{minipage}[t]{0.33\linewidth}
\centering
\includegraphics[width=2.5in]{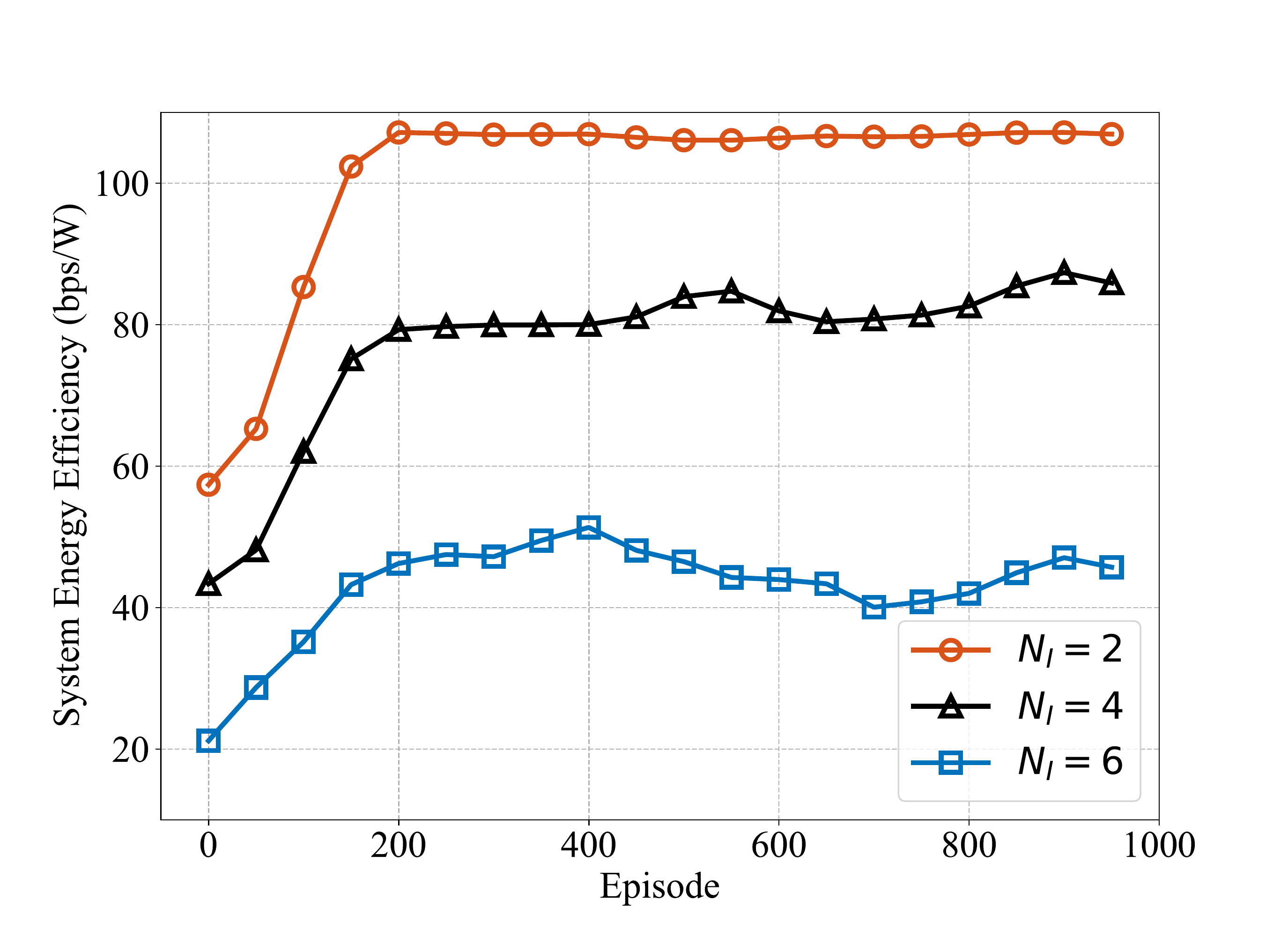}
\end{minipage}%
}%
\subfigure[]{
\begin{minipage}[t]{0.33\linewidth}
\centering
\includegraphics[width=2.5in]{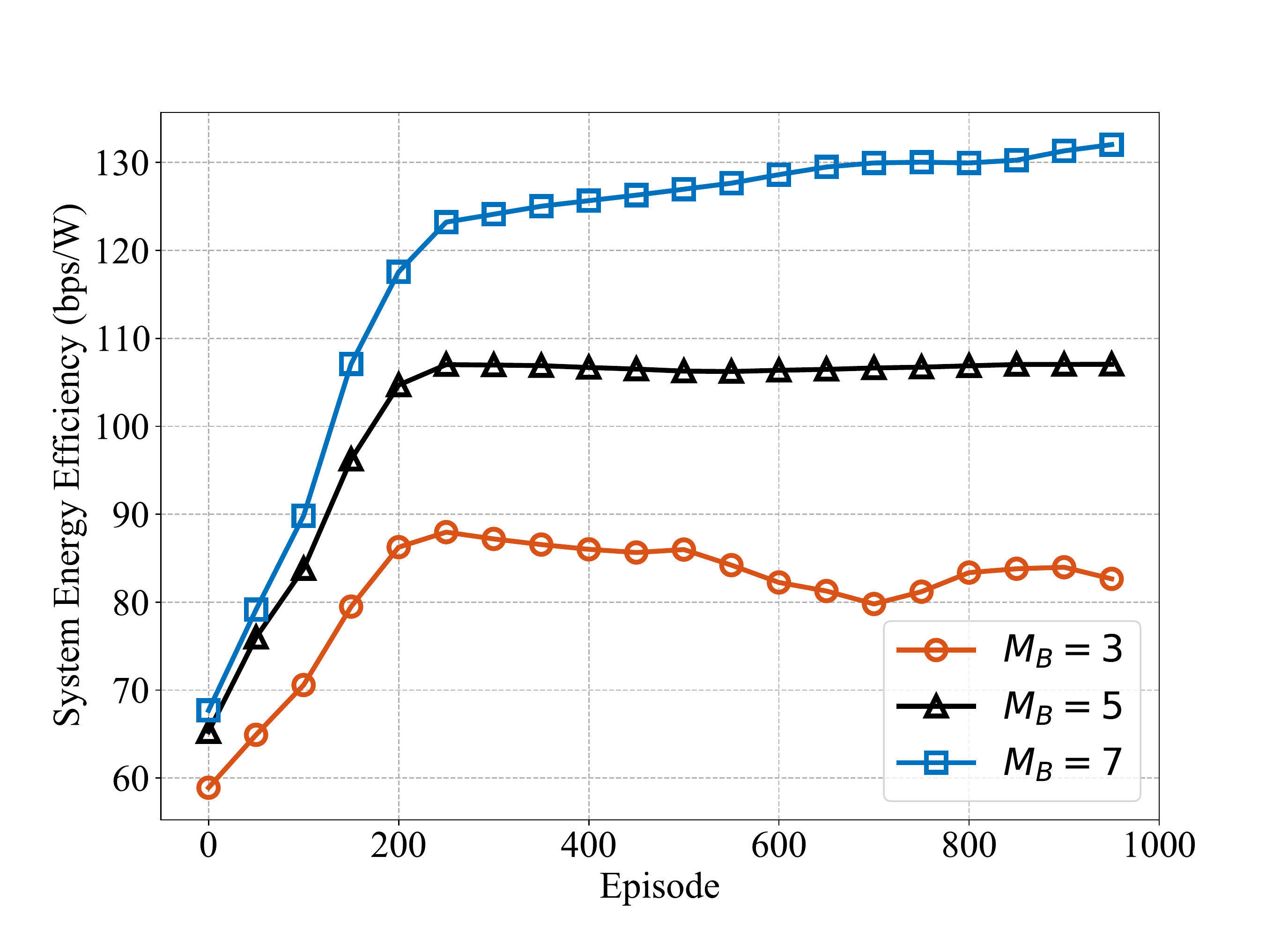}
\end{minipage}%
}%

\caption{System energy efficiency versus episode: a) over the different number of IRS reflection elements; b) over the different number of IoT devices; c) over the different number of BS antennas.}
\label{EE-epesode-system}
\end{figure*}


\begin{figure*}[htbp]
\centering
\subfigure[]{
\begin{minipage}[t]{0.33\linewidth}
\centering
\includegraphics[width=2.5in]{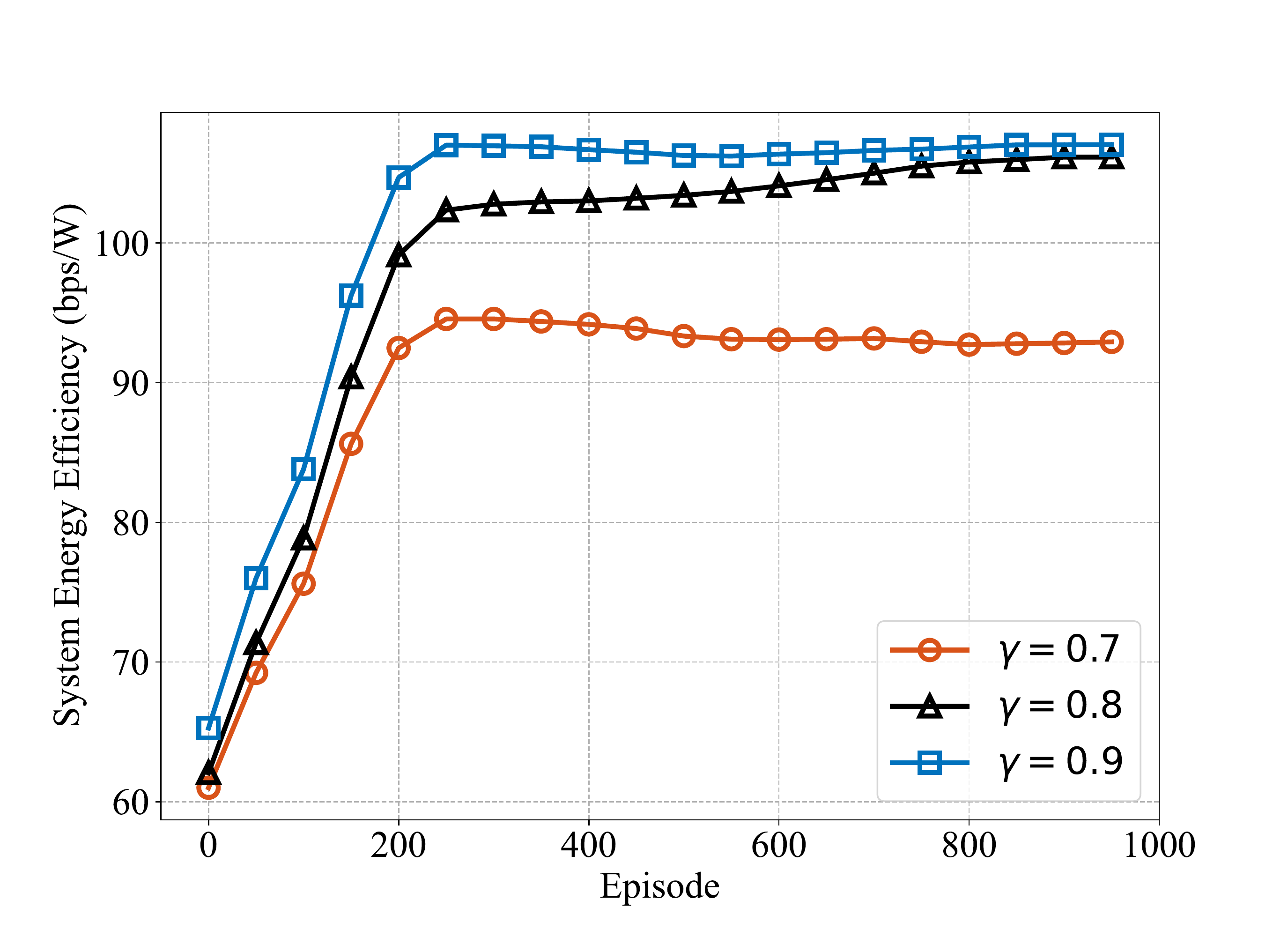}
\end{minipage}%
}%
\subfigure[]{
\begin{minipage}[t]{0.33\linewidth}
\centering
\includegraphics[width=2.5in]{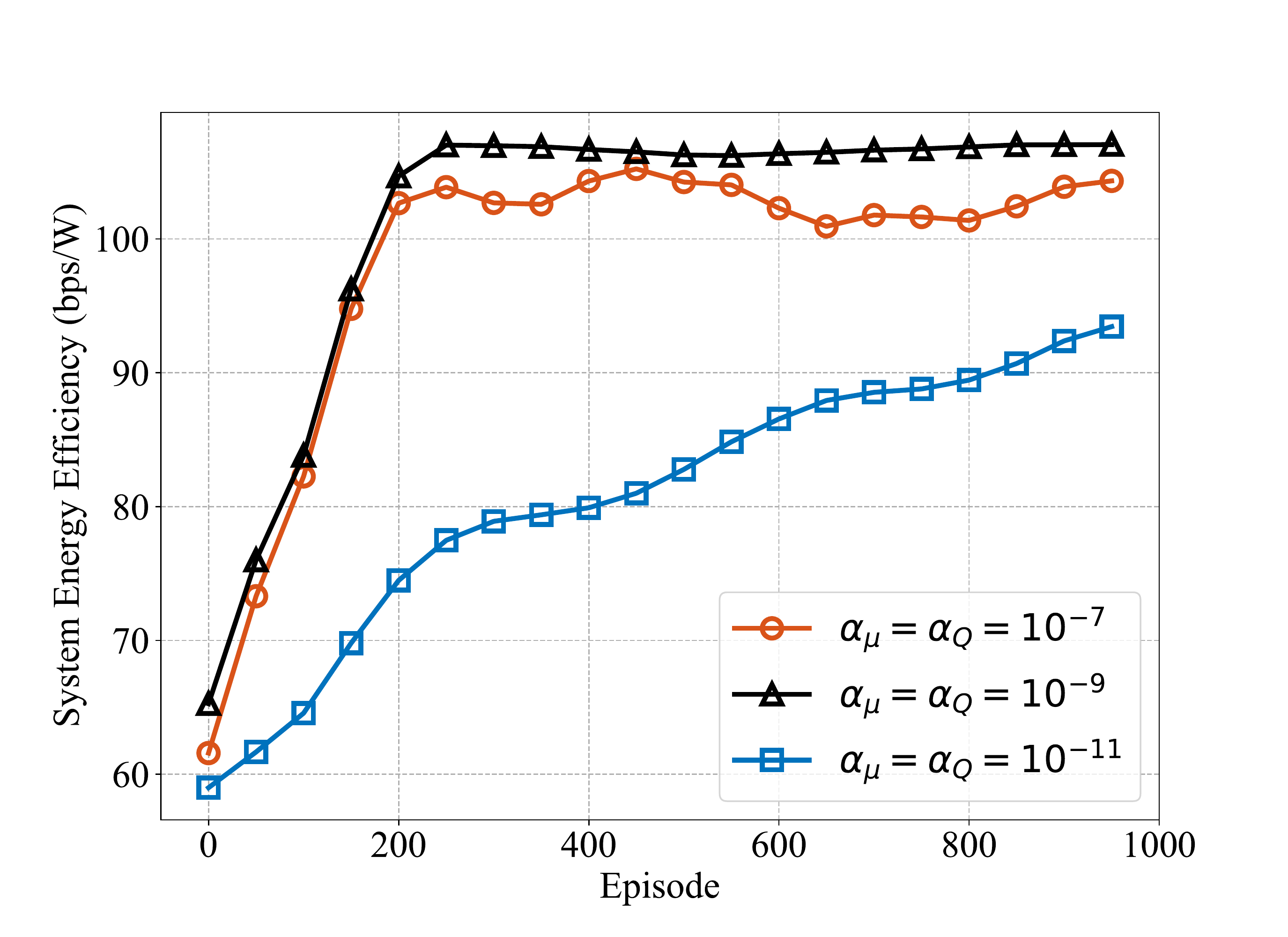}
\end{minipage}%
}%
\subfigure[]{
\begin{minipage}[t]{0.33\linewidth}
\centering
\includegraphics[width=2.5in]{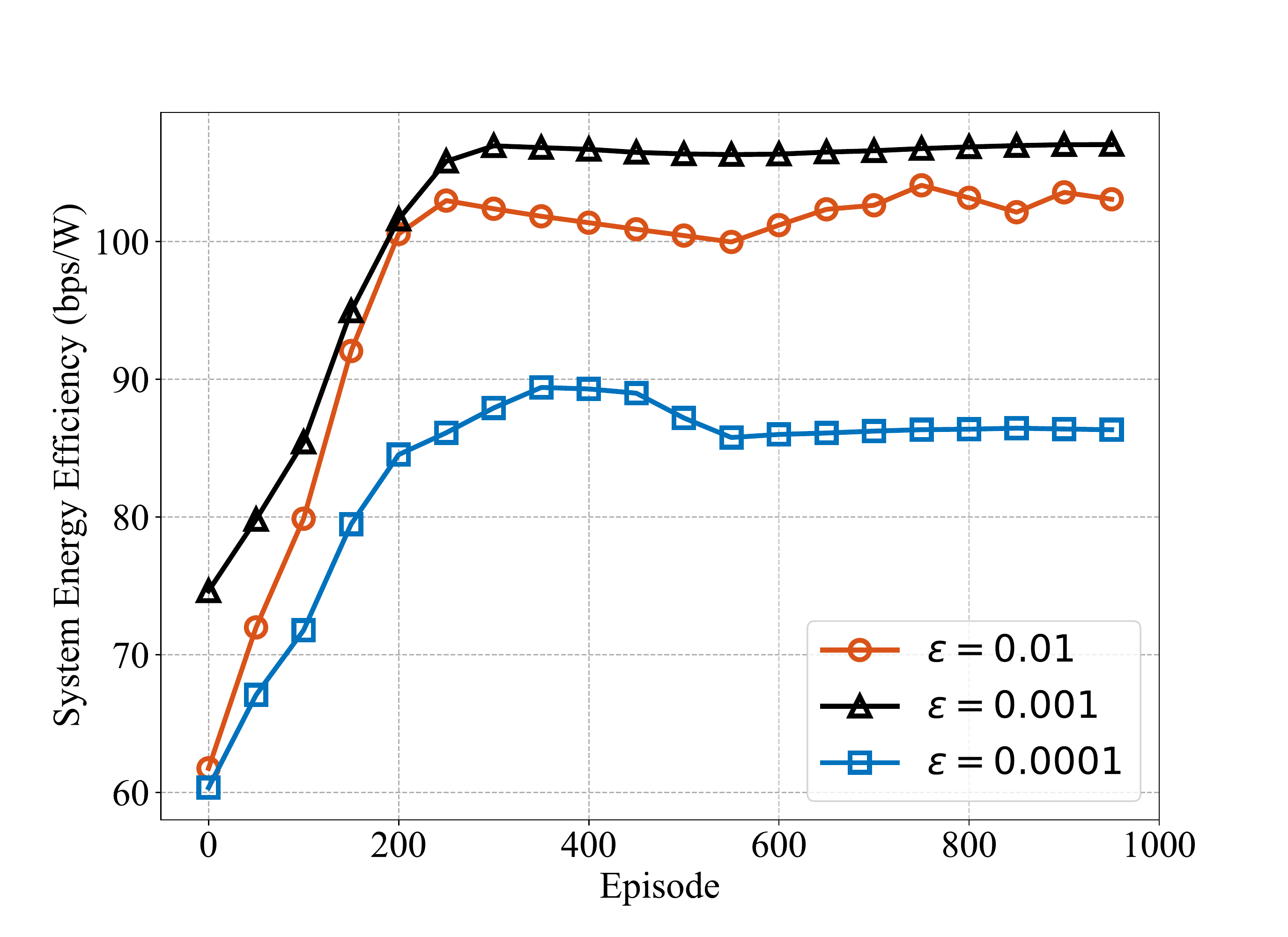}
\end{minipage}%
}%

\caption{DRL-driven system energy efficiency versus episode: a) over the various discount factor of DRL; b) over the various learning factors of DRL; c) over the various update rate of DRL.}
\label{EE-epesode-DRL}
\end{figure*}

\subsection{Comparison Analysis with Baseline Strategy}
\label{other}
To demonstrate the effectiveness of our proposed 6G IoT communication mechanism, we compare the proposed DRL and IRS empowered mechanism with other strategy in terms of the utility (i.e. energy efficiency) of the whole network, which are described as follows:
\begin{itemize}
  \item With IRS: The mechanism with IRS refers to our proposed AI and IRS empowered strategy, which optimizes the IRS reflection phase, IoT power, and BS detection matrix based on the proposed DRL-driven resource control scheme.
  \item Without IRS: The mechanism without IRS is the baseline strategy we compare with, which is the common smart resource allocation strategy for data transmission without the assistance of IRS. In the baseline strategy, DRL is used to optimize the settings of BS and IoT devices.
\end{itemize}

Experimental results of the comparison between the mechanism with IRS and without IRS are shown in Fig. \ref{EE-BS} and Fig. \ref{EE-IoT}. From the results in Fig. \ref{EE-IoT} and \ref{EE-BS}, we can see that along with the rising number of IoT devices, the system energy efficiency decreases, while energy efficiency increase as the rising of BS antennas. These phenomena are consistent with the results shown in Fig. \ref{EE-epesode-system} (b) and (c), and will be analyzed in detail in the next subsection. More importantly, on all different numbers of IoT devices and BS antennas, our proposed AI and IRS-aided 6G IoT communication system achieves better performance than that without IRS.

\subsection{Impact of Parameter Settings on Our Mechanism}
\label{results}

This subsection investigates the impact of parameter settings on our mechanism. The 6G IoT system energy efficiency versus the iterations under the various number of system devices is presented in Fig. \ref{EE-epesode-system}. We can obtain that along with the rising episode, the system energy efficiency increases for all curves and converges to certain values. The system energy efficiency versus episode over the different number of IRS-reflection elements is shown in Fig. \ref{EE-epesode-system} (a), where the number of IRS-reflection elements is set as $M_R=10, 20, $ and $30$. The curve of $M_R=30$ lies above the curve of $M_R=20$ and $M_R=10$. That means deploying more IRS elements within a certain extent can gain more system benefits, i.e. faster convergence and higher reward. That is because the IRS has the capability to improve signal propagation with low power consumption, so adding IRS-reflection elements can improve system performance.

We also explore the impact of the number of IoT devices on system performance and present the simulation results in Fig. \ref{EE-epesode-system} (b). As shown in this subfigure, more IoT users need to send data in the cell results in lower system energy efficiency. This results from that under the definite total amount of network resources (e.g. BS antennas, IRS elements, and etc.), the 6G IoT communication system needs to supports more users and transmits more data at the same time with the increasing of IoT users. Thus, the system energy efficiency decreases as more IoT users involve in.

The system energy efficiency under the different number of BS antennas is investigated in Fig. \ref{EE-epesode-system} (c), where the number of BS antennas are set as $M_B=3,5$ and $7$. From these experimental results, we can obtain that increasing the number of BS antennas can improve the overall energy efficiency of the system. As an important resource of the communication system, the BS antennas support signal transmission and reception. Thus, increasing the number of BS antennas within a certain extent can provide better services and system performance.

The experimental results of the 6G IoT communication system energy efficiency versus episode over various DRL parameter settings are shown in Fig. \ref{EE-epesode-DRL}. We investigate the impact of the discount factor for the proposed DRL and IRS-driven system in Fig. \ref{EE-epesode-DRL} (a). The energy efficiency of different $\gamma$ increases and converges to certain values as the episode rises. The curve with discount factor $\gamma=0.9$ obtains better performance than $\gamma=0.7$ and $0.8$. The energy efficiency of the proposed 6G IoT system for various learning factors of actor-critic networks is depicted in Fig. \ref{EE-epesode-DRL} (b). The curve with learning factors $\alpha_{\mu}=\alpha_Q=10^{-9}$ achieves the best performance, which converges at a faster rate and higher level. In Fig. \ref{EE-epesode-DRL} (c), we explore the energy efficiency over various update rate of the proposed DRL scheme. Compared with too large or too small update rate, i.e. $\varepsilon=0.01$ or $\varepsilon=0.0001$, the proposed DRL-based resource control and allocation scheme with $\varepsilon=0.001$ achieves the higher system performance.

\section{Conclusion}
\label{sec:conclusion}
In this article, we propose a DRL and IRS empowered energy-efficiency communication system for 6G IoT. We first propose an intelligent and efficient communication architecture, where an IRS-aided data transmission mechanism and a DRL-based resource management mechanism are designed. Then, an optimization model is formulated to maximize the system energy efficiency under given delay requirements. Next, based on network and channel status, a DRL-driven network resource control and allocation scheme is designed to solve the formulated optimization model and improve system performance. Finally, experimental results verify the effectiveness of our proposed energy-efficiency communication system for 6G IoT. Future work is to propose a latency-minimization transmission system for 6G IoT under given energy constraints.


%

%

\ifCLASSOPTIONcompsoc
  \section*{Acknowledgments}
\else
  \section*{Acknowledgment}
\fi

This work is sponsored by the National Natural Science Foundation of China (Grant No.  61972255).

\ifCLASSOPTIONcaptionsoff
  \newpage
\fi



%
%
%

\bibliographystyle{IEEEtran}
\bibliography{IEEEabrv, ./bib/mylib}

\end{document}